\documentclass[11pt]{cernrep}
\usepackage{graphicx}

\newcommand{\Et}{{E_t}}

\newcommand{\Ptgp}{\Et^{\gamma,\pi^0}}
\newcommand{\Ptall}{\Et^{\gamma,\pi^0,\eta,K}}

\newcommand{\lt}{\!<\!}
\newcommand{\gt}{\!>\!}

\newcommand{\wrt}{with respect to }
\newcommand{\gpj}{~``$\gamma+Jet$''~}

\begin{document}
\thispagestyle{empty}
 
\vskip-5mm
 
\begin{center}
{\Large JOINT INSTITUTE FOR NUCLEAR RESEARCH}
\end{center}
 
\vskip10mm
 
\begin{flushright}
JINR Communication \\
E2-2001-259 \\
hep-ex/0108051
\end{flushright}
 
\vspace*{3cm}
 
\begin{center}
\noindent
{\Large{
\bfseries Separation of a single photon and products of %\\[0pt]
the $\pi^0,\eta, K^0_s$ meson neutral decay channels in %\\[4pt]
 the CMS electromagnetic calorimeter using neural network.}}\\[10mm]
{\large D.V.~Bandourin$^{1}$, N.B.~Skachkov$^{2}$}
 
\vskip 0mm
 
{\small
{\it
E-mail: (1) dmv@cv.jinr.ru, (2) skachkov@cv.jinr.ru}}\\[3mm]
%$\dag$ 
\large \it Laboratory of Nuclear Problems \\
\end{center}
 
\vskip 11mm
\begin{center}
\begin{minipage}{150mm}
\centerline{\bf Abstract} 
~\\[-1pt]
\noindent
The artificial neural network approach is used  for  separation of
signals from a single photon $\gamma$ and products of the
$\pi^0,\eta, K^0_s$ meson neutral decay channels on the basis of the data
from the CMS electromagnetic calorimeter alone. Rejection values for 
the three types of mesons as a function of single photon selection
efficiencies are obtained for two Barrel and one Endcap pseudorapidity
regions and initial $\Et$ of 20, 40, 60 and 100 $GeV$.
\end{minipage}
\end{center}       

\newpage
 
\setcounter{page}{1}    
%===========================================================
%
\section{Introduction.}
% #1
%===========================================================

In our previous papers \cite{SF}, \cite{BKS_P1} it was proposed to use the direct
photon production process based at the partonic level on the
Compton-like QCD subprocess $q g\to q+\gamma$
for extracting the gluon distribution function $f^g(x,Q^2)$ in a proton at the LHC. 
One of the main background sources, as  was established in \cite{BKS_P5}, 
%to the direct photon signal comes from 
is the photons produced in neutral decay channels 
of $\pi^0,\eta$ and $K^0_s$ mesons
\footnote{Along with bremsstrahlung photons produced from a quark
in the fundamental $2\to 2$ QCD subprocess (see \cite{SF},
\cite{BKS_P5}).}.
So, to obtain a clean sample of events for gluon distribution function determination
with a low background contamination it is necessary to discriminate between the direct photon 
signal and the signal from the photons produced in the neutral decay channels of $\pi^0,\eta$ 
and $K^0_s$ mesons.

Among other physical processes in which one needs to separate a single photon from the
background photons one can note the $H^0\to \gamma \gamma$ decay. Obtaining clean
signals from $\gamma$'s in this process would enhance an accuracy of the
Higgs boson mass determination.

%===========================================================
%
\section{Data simulation.}
% #2
%===========================================================

There is a number of the CMS publications on  photon and neutral pion
discrimination (see \cite{Bar1}, \cite{Bar2}, \cite{Bor}).

Information from the electromagnetic calorimeter (ECAL) crystal cells alone is used 
in this paper to extract a photon signal. 
%for analysis both in the Barrel and Endcap regions. 
ECAL cells were analyzed after performing the digitization procedure
\footnote{Special thanks to A.~Nikitenko for his help with digitization routines.}.

The GEANT-based full detector simulation package CMSIM (version 121) for CMS \cite{CMSIM} 
was used.
We carried out a set of simulation runs including: (a) four particle types
$\gamma$ and $\pi^0,\eta, K^0_s$ mesons, which were forced to decay only via neutral channels
 (see Table~\ref{tab:PDG} based on the PDG data \cite{PDG});
(b) five $\Et$ values 20, 40, 60, 100 and 200 $GeV$;
(c) three pseudorapidity $\eta$ intervals $|\eta|\lt0.4$, $1.0\lt\eta\lt1.4$ 
(two Barrel regions) and $1.6\lt\eta\lt2.4$ (Endcap region).

About 4000 single particle events were generated for the CMSIM simulation of each type.
The information from the $5\times 5$ ECAL crystal cells window (ECAL tower) with the
most energetic cell in the center was used in the subsequent analysis based on
the artificial neural network (ANN) approach. The application of a software-implemented
neural network for pattern recognition and triggering tasks is well known.
This study was carried out with the JETNET 3.0 package
\footnote{It is available via {\it anonymous} {\bf ftp} from {\bf thep.lu.se} or from
{\bf freehep.scri.fsu.edu}.}
developed at CERN and the University of Lund \cite{JN}.\\[-17pt]
\def\baselinestretch{0.95}
\begin{table}[h]
\begin{center}
\caption{Decay modes of $\pi^0, \eta$ and $K^0_s$ mesons.}
\label{tab:PDG}
\vskip0.2cm
\begin{tabular}{|c|c|c|}                  \hline
Particle & Br.($\%$)& Decay mode  \\\hline
$\pi^0$ & 98.8 & $\gamma\gamma$    \\
        &  1.2 & $\gamma e^+e^-$    \\\cline{1-3}
$\eta$  & 39.3 & $\gamma\gamma$    \\
        & 32.2 & $\pi^0\pi^0\pi^0$ \\
        & 23.0 & $\pi^+\pi^-\pi^0$ \\
        & 4.8  & $\pi^+\pi^-\gamma$ \\\cline{1-3}
$K^0_s$ & 68.6 & $\pi^+\pi^-$ \\
        & 31.4 & $\pi^0\pi^0$ \\\cline{1-3}
$\omega$& 88.8 & $\pi^+\pi^-\pi^0$ \\
        &  8.5 & $\pi^0\gamma$ \\\cline{1-3}
\end{tabular}
%\vskip-10mm
\end{center}
\end{table}
\def\baselinestretch{1.0}

%===========================================================
%
\section{Neural network architecture and input data.}
% #3
%===========================================================

The feed forward ANN with 11 input and 5 hidden nodes in one
hidden layer and with binary output was chosen for analysis, i.e. with the 11 -- 5 -- 1 architecture. 
``Feed forward'' implies that information can
only flow in one direction (from input to output) and the ANN
output directly determines the probability that an event characterized
by some input pattern vector $X(x_1,x_2,...,x_n)$ ($n=11$ in our
case) is from the signal class.

The following data received from the ECAL tower were put on the 11 network input nodes
($0th$ layer):\\
\hspace*{5mm} $1-9$: data from the first nine crystal cells ordered with respect to energy $E$:
$E$ of the leading cell was assigned to the $1st$ input node,
$E$ of the next-to-leading cell to the $2nd$ input node and so on. \\
\hspace*{5mm}  $10, 11$: Two ``width'' variables defined as:\\[-7pt]
\begin{equation}
\eta_{w}=\frac{\sum\limits_{i=1}^{25} E_i\left(\eta_i - \eta_{cog}\right)^2}
{\sum\limits_{i=1}^{25} E^i},
\quad
\phi_{w}=\frac{\sum\limits_{i=1}^{25} E_i\left(\phi_i - \phi_{cog}\right)^2}
{\sum\limits_{i=1}^{25} E^i}.
\label{eq:COG}
\end{equation}
Here $(\eta_{cog},\phi_{cog})$ are the coordinates of the center of
gravity of the ECAL tower considered. It was established that
variation of the crystal cell number from 7 to 12 practically does not change the
network performance.

To ensure convergence and stability the total number of training
patterns must be significantly (at least $20-30$ times) larger than
the number of independent parameters of the network given by
formula:\\[-20pt]
\begin{eqnarray}
N_{ind}=(N_{in}+N_{on})\cdot N_{hn} + N_{ht} + N_{ot}
\label{eq:n_ind}
\end{eqnarray}
where $N_{in}$ is the number of input nodes; $N_{on}$ is the number of output nodes;
$N_{hn}$ is the number of nodes in a hidden single layer; $N_{ht}$ is the number of thresholds in
a hidden single layer; $N_{ot}$ is the number of output thresholds (here $N_{on}=N_{ot}=1$).
So, for our 11 -- 5 -- 1 architecture we get $N_{ind}=66$.

%====================================================================
\section{Training and testing of ANN.}
% #4
%====================================================================
There are two stages in neural network (NN) analysis. The first is
the training/learning of the network with samples of signal and background
events and the second is testing stage using independent data
sets. Learning is the process of adjusting $N_{ind}$ independent parameters
in formula (\ref{eq:n_ind}). The training  starts with random weights values.
After feeding the training input vector (see 11 variables in the previous section), the NN output
$O^{(p)}$  is calculated for every training pattern $p$ and
compared with the target value $t^{(p)}$, which is 1 for 
single photon and 0 otherwise. After $N_p$ events are presented to
the network, the weights are updated by minimization of the mean
squared error function $E$ averaged over the number of training
patterns:\\[-19pt]
\begin{eqnarray}
E=\frac{1}{2N_p} \sum\limits_{p=1}^{N_p} (O^{(p)} - t^{(p)})^2,
\label{eq:err}
\end{eqnarray}
\vskip-3mm
\noindent
where $O^{(p)}$ is the output value for a pattern $p$,
$t^{(p)}$ is the training target value for this pattern $p$,
$N_p$ is the number of patterns (events) in the training sample per update of weights
(here $N_p$ is equal to 10, the JETNET default value).

\begin{figure}[htbp]
\vskip-28mm
\hspace*{-10mm} \includegraphics[width=16cm,height=7cm,angle=0]{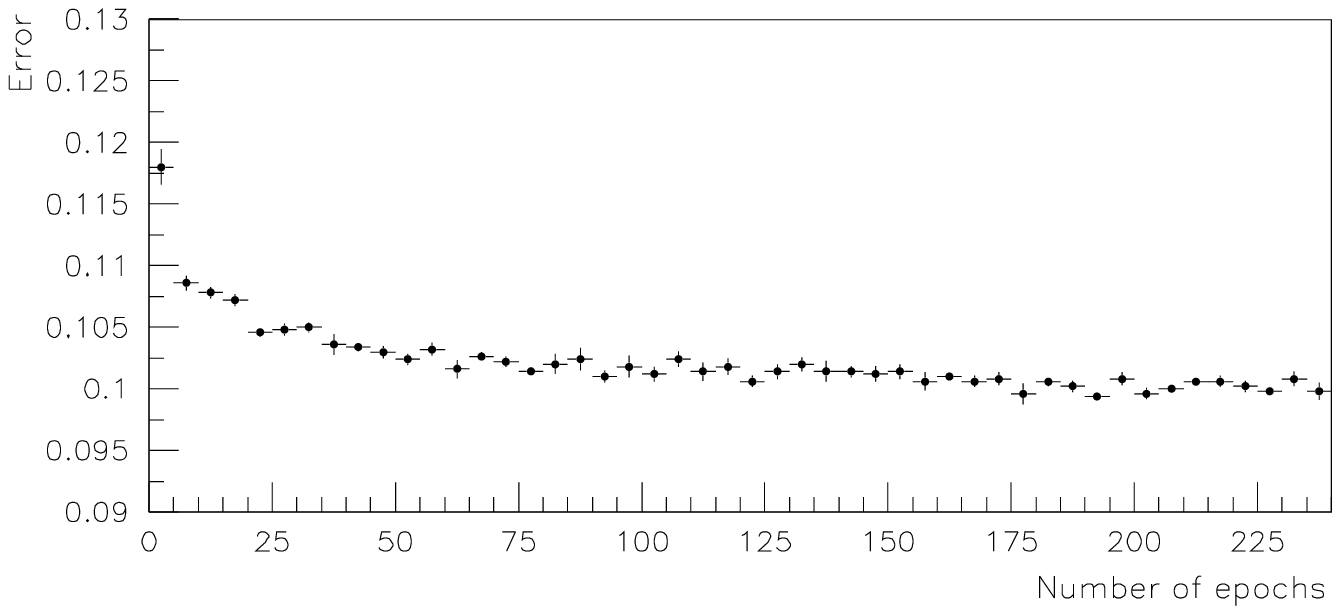}
\vskip-12mm
  \caption{A dependence of a mean error per epoch on the epoch number during the training procedure
($|\eta|\lt0.4$, $\Et=40 ~~GeV$).}
  \label{fig:err}
%\end{figure}
%\begin{figure}[htbp]
\vskip-10mm
\hspace*{-10mm} \includegraphics[width=16cm,height=7.5cm,angle=0]{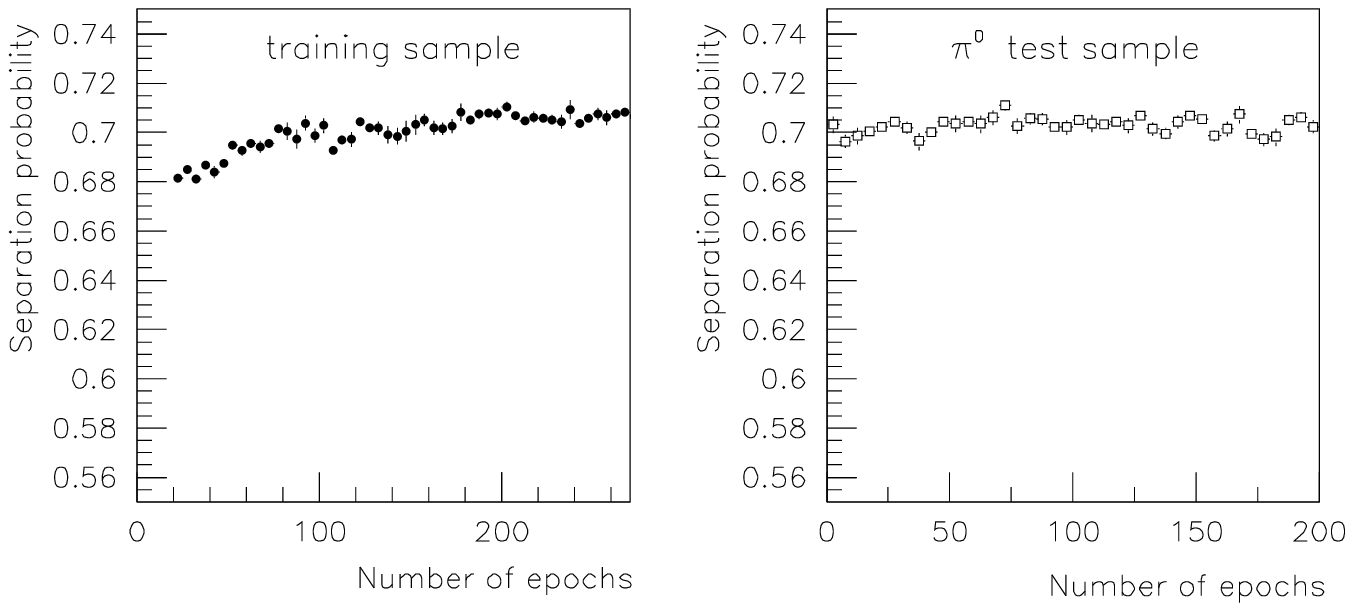}
\vskip-10mm
  \label{fig:prof_p}
%\end{figure}
%\begin{figure}[htbp]
%\begin{flushleft}
\vskip-9mm
\hspace*{-10mm} \includegraphics[width=9cm,height=7.5cm,angle=0]{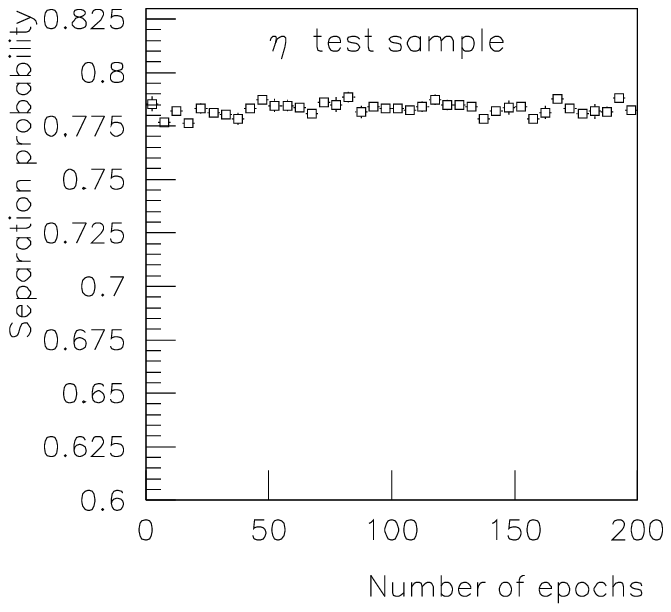}
\vskip-10mm
  \caption{A dependence of a separation probability on the epoch number for training $\pi^0$ sample
and test samples of $\pi^0, \eta, K^0_s$ mesons ($|\eta|\lt0.4$, $\Et=40 ~~GeV$) for the network 
output threshold $O_{thr}=0.5$.}
  \label{fig:prof_e}
%\end{flushleft}
%\begin{flushright}
\vskip-79mm
\hspace*{60mm} \includegraphics[width=9cm,height=7.5cm,angle=0]{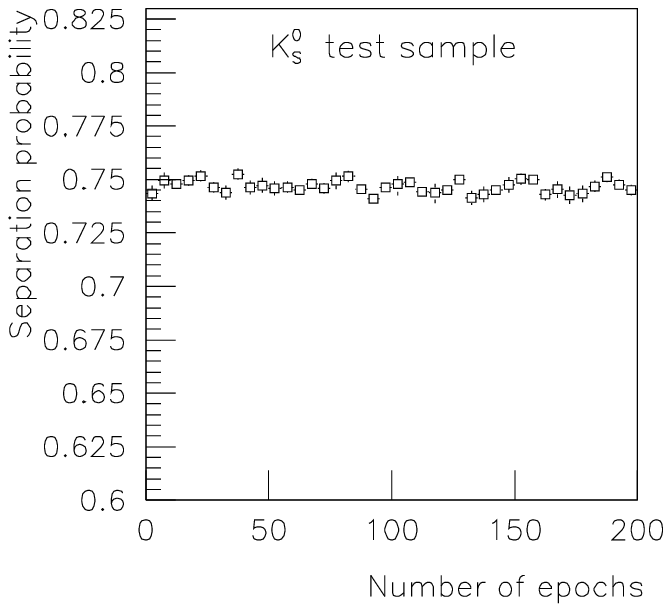}
\vskip-10mm
%  \caption{}
  \label{fig:prof_k}
\nonumber
%\end{flushright}
\end{figure}

This error is decreased during the network training procedure. Its
behavior during training is shown in Fig.~\ref{fig:err} (we see
that it drops as 0.117 $\to$ 0.096). Here ``Number of epochs'' is the
number of training sessions (equal to 200 here). 
For each epoch the percentage of correctly classified events/patterns
is calculated with respect to the neural network threshold
$O_{thr}=0.5$, classifying the input as a ``$\gamma$-event'' if the NN
output $O\gt0.5$ and as a ``background ($\pi^0,\eta,K^0_s$) event'' if
the NN output $O\lt0.5$. Below we shall call this criterion the ``0.5-criterion''
%
%\footnote{It is worth emphasizing that for practical applications various
\footnote{It should be noted that for practical applications various
$O_{thr}$ values can be used (see Section 5).}.
About 3000 signal  events/patterns (containing the ECAL data from single photons) and each type of the
background  events/patterns (containing the ECAL data from the multiphoton $\pi^0,\eta,K^0_s$ meson
decays) were chosen for training stage, i.e. more than 90 patterns per weight.

\begin{center}
\begin{figure}[htbp]
\vskip-1mm
\hspace*{-4mm} \includegraphics[width=15cm,height=17cm,angle=0]{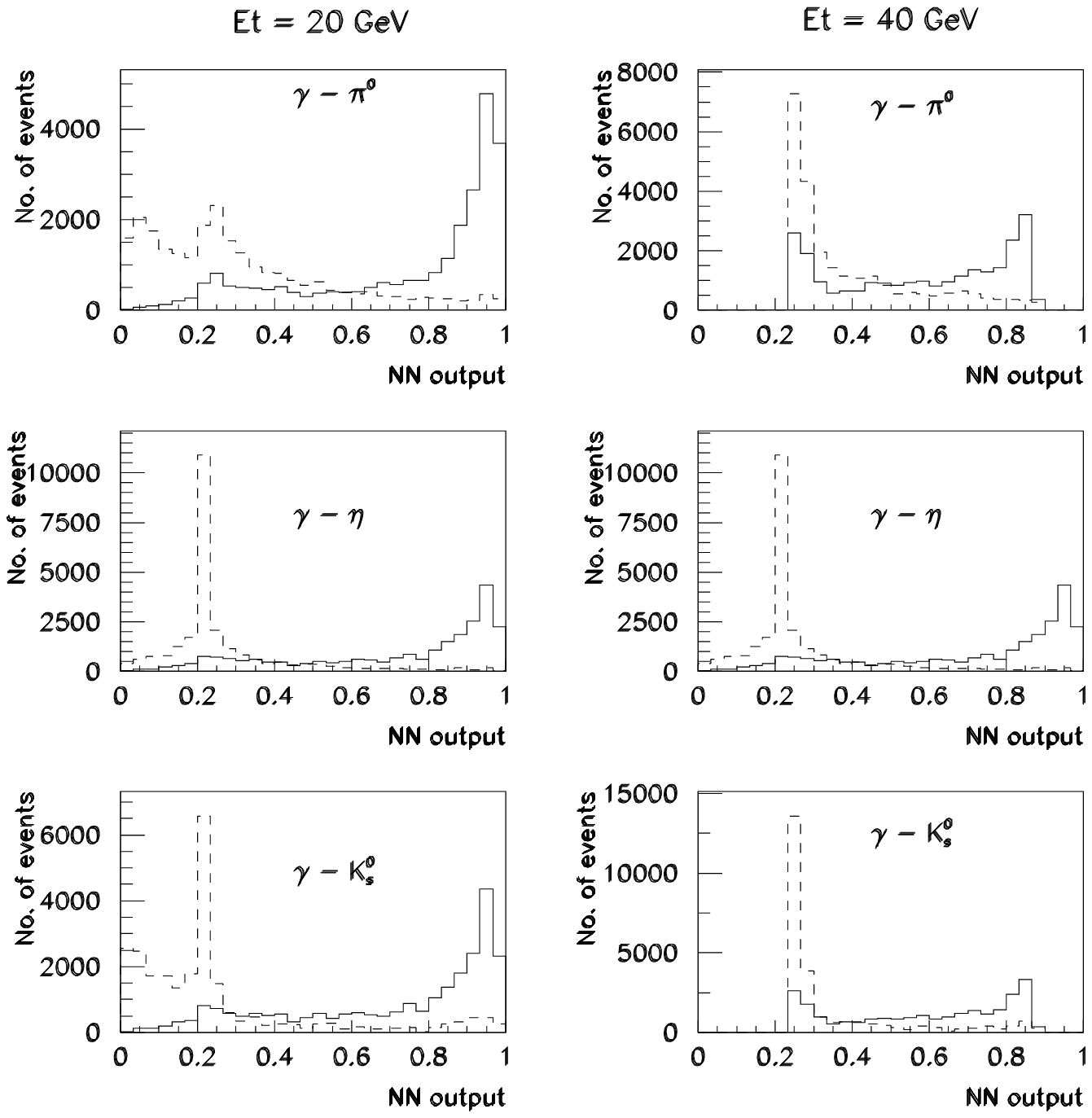}
\vskip-1mm
  \caption{Neural network output for the test samples of $\pi^0, \eta, K^0_s$ mesons (dotted lines)
and photon (solid line) ($|\eta|<0.4$, $\Et = 20, 40 ~~GeV$).}
  \label{fig:nn_out1}
\end{figure}
\end{center}

\begin{center}
\begin{figure}[htbp]
\vskip-1mm
\hspace*{-4mm} \includegraphics[width=15cm,height=17cm,angle=0]{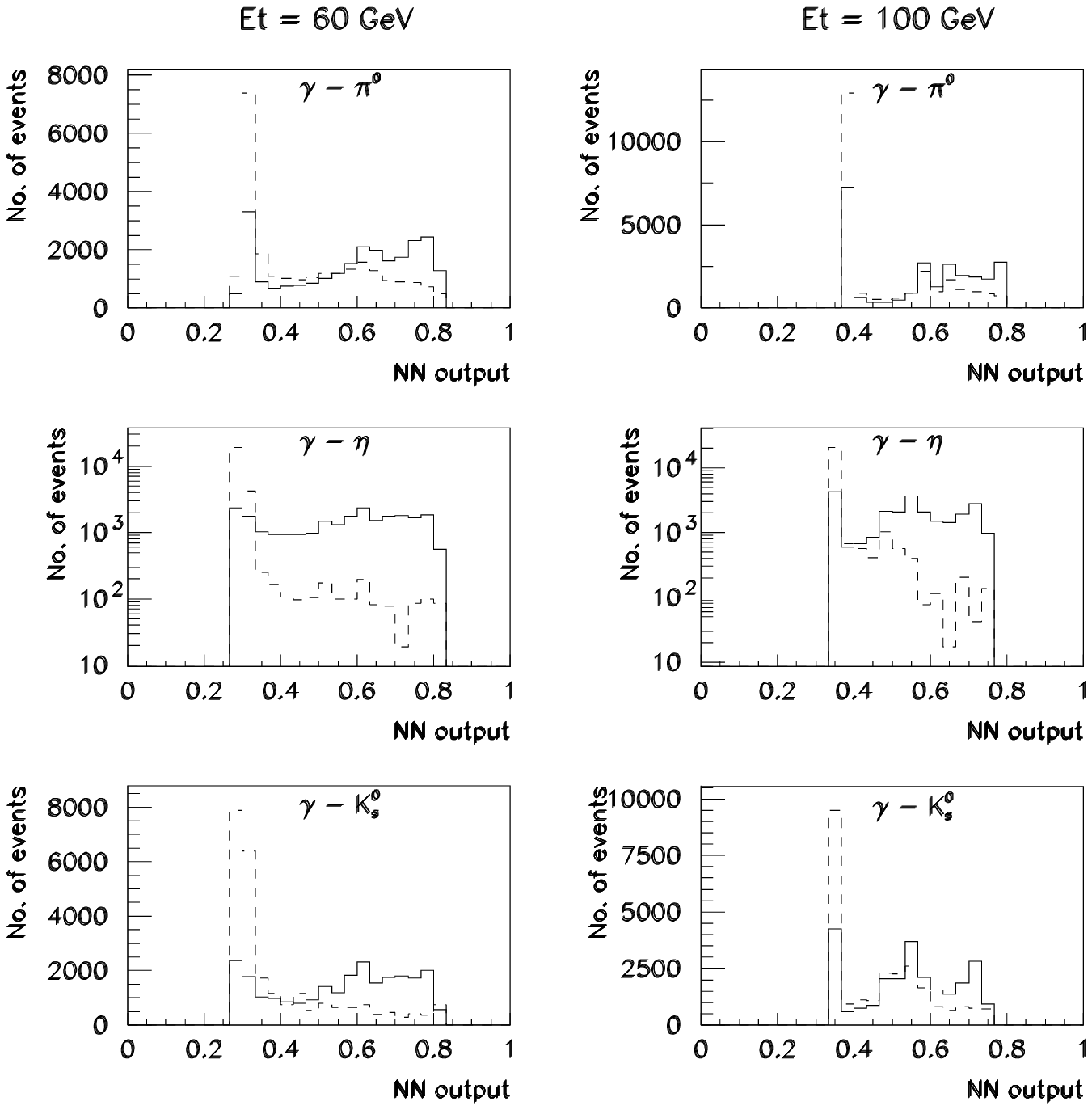}
\vskip-1mm
  \caption{Neural network output for the test samples of $\pi^0, \eta, K^0_s$ mesons (dotted lines)
and photon (solid line) ($|\eta|<0.4$, $\Et = 60, 100 ~~GeV$).}
  \label{fig:nn_out2}
\end{figure}
\end{center}

%Testing
After the network was trained, a test procedure was implemented in which the events
not used in the training were passed through the network.
% eta and K0S discrimination
The sets of weights obtained after the neural network training with the $\gamma/\pi^0$
samples were written to a file for every $\Et$ and pseudorapidity interval.
Then, the {\it the same set of weights} read from the corresponding file was applied
to test sets of the $\pi^0,\eta,K^0_s$ events which the network had never seen before
to find a test (generalization) performance with respect to every type of input event set.
2000 signal and background events (about 1000 of each sort)

\noindent
 were used at the generalization stage.
The output provided for each event can be considered as a probability that this event 
is either from the signal or the background
sample. If the training is done correctly, the probability for an event to be
a signal is high if the output $O$ is close to 1. And conversely, if the output $O$ is
close to 0, it is more likely to be a background event. The
network performance with respect to the ``0.5-criterion'' as a
function of the training and test epoch number (for each type of
background) is presented in Fig.~\ref{fig:prof_e}.
One can see from the  ``training sample'' plot (upper left corner) 
that the neural network performance becomes
stable starting with the epoch number 180--200. In
Figs.~\ref{fig:nn_out1} and \ref{fig:nn_out2} we show the neural
network output for the test samples of $\pi^0, \eta, K^0_s$ mesons
with $\Et = 20, 40, 60, 100 ~~GeV$ (the $|\eta|\lt0.4$ interval was
taken as an example). One can see that the range of the network
output values becomes narrower with growing $\Et$ and, consequently,
 signal and background event classes become less distinguishable.

The Manhattan algorithm for weight updating was used at the training stage.
In Table~\ref{tab:algo} we compare it with other popular in high energy physics
updating algorithms with varying learning rate $\eta$ (Backpropagation, Langevin)  
and noise term $\sigma$ (Langevin) for a case of photons in two Barrel regions with $\Et=40~ GeV$. 
%(the first column of each algorithm corresponds to the default values in the JETNET package).
\\[-27pt]
\begin{table}[h]
\begin{center}
%\vskip5mm
\caption{\footnotesize A dependence of the separation probability ($\%$) using
 ``0.5-criterion'' on the method. $\Et=40 ~~GeV$, Barrel region.}
\vskip1mm
\begin{tabular}{||c||c|c|c|c||c|c|c||}                  \hline \hline
\label{tab:algo} Method
&\multicolumn{4}{c||}{Backpropagation}&\multicolumn{3}{c||}{Langevin}
\\\cline{1-8}
Parameters: $\eta$    &1.0 & 0.1 & 0.01 & 0.001 &1.0 &0.1 -- 0.01 & 0.01    \\\cline{2-8}
\hspace*{21mm}$\sigma$& -- & --  &  --  &  --   &0.01&0.01     &0.001 \\\hline \hline
$|\eta|\lt0.4$  & 51 & 65 & 67 & 65 & 50 & 60 & 66   \\\hline
$1.0\lt\eta\lt1.4$& 50 & 64 & 65 & 65 & 50 & 60 & 64   \\\hline
\end{tabular}
\end{center}
\vskip-6mm
\end{table}

\noindent 
We shall see that even the best results obtained with other
algorithms (e.g. for  $|\eta|\lt0.4$ we get $67\%$ for the Backpropagation and 
$66\%$ for Langevin algorithms) 
are by $3-4\%$ worse than the results obtained with the Manhattan algorithm:
$70\%$ for $|\eta|\lt0.4$  and $68\%$ for $1.0\lt\eta\lt1.4$
(see Tables \ref{tab1} and \ref{tab2} of the next section).

%===============================================================================
%
\section{Description of the results.}
% #5
%===============================================================================

The discrimination powers for various  types of test event samples \wrt the middle point
criterion (i.e. $O_{thr}=0.5$) are presented in Tables~\ref{tab1} -- \ref{tab3} 
for three pseudorapidity intervals and four $\Et$ values.

We see first that for the both Barrel regions the $\gamma/\pi^0$ (and $\gamma/K^0_s$)
separation efficiencies drop as $75-79\%$ ($79-82\%$) to $60-61\%$ ($56-63\%$) while
for $\gamma/\eta$ they practically do not change and remain as large as
$80\%$. All separation efficiencies substantially decrease  when we come to the Endcap region
(Table ~\ref{tab3}).

%\\[-7mm]
\begin{table}[h]
\begin{center}
\caption{The separation probability ($\%$) using ``0.5-criterion''.
$|\eta|\lt 0.4$.}
\vskip.1cm
\begin{tabular}{||c||c|c|c|c||}                  \hline \hline
\label{tab1}
Particle& \multicolumn{4}{c|}{$\Ptall$ value ($GeV)$} \\\cline{2-5}
 type  & 20 & 40 & 60 & 100   \\\hline \hline
$\pi^0$& 79 & 70 & 64 & 60   \\\hline
$\eta$ & 83(87) & 79(88) & 80(88) & 80(84) \\\hline
$K^0_s$& 82(84) & 75(79) & 71(73) & 63(66)  \\\hline
\hline
\end{tabular}
\end{center}
\vskip-6mm
%\hspace*{17mm}\footnotesize{The error is of order of $1.5-2\%$ for all numbers}
%\end{table}
\vskip-5mm
%\begin{table}[h]
\begin{center}
\caption{The separation probability ($\%$) using ``0.5-criterion''. 
$1.0\lt\eta\lt1.4$.}
\vskip.1cm
\begin{tabular}{||c||c|c|c|c||}                  \hline \hline
\label{tab2}
Particle& \multicolumn{4}{c|}{$\Ptall$ value ($GeV)$} \\\cline{2-5}
 type  & 20 & 40 & 60 & 100   \\\hline \hline
$\pi^0$& 75 & 68 & 63 & 61 \\\hline
$\eta$ & 79(83) & 77(84) & 78(84) & 78(79) \\\hline
$K^0_s$& 79(83) & 69(75) & 66(70) & 56(59)  \\\hline
\hline
\end{tabular}
\end{center}
\vskip-6mm
%\hspace*{17mm}\footnotesize{The error is of order of $1.5-2\%$ for all numbers}
%\end{table}
\vskip-5mm
%\begin{table}[h]
\begin{center}
\caption{The separation probability ($\%$) using ``0.5-criterion''.
$1.6\lt\eta\lt2.4$.}
\vskip.1cm
\begin{tabular}{||c||c|c|c|c||}                  \hline \hline
\label{tab3}
Particle& \multicolumn{4}{c|}{$\Ptall$ value ($GeV)$} \\\cline{2-5}
 type  & 20 & 40 & 60 & 100  \\\hline \hline
$\pi^0$& 63 & 59 & 56 & 54   \\\hline
$\eta$ & 72(77) & 74(76) & 66(68) & 63(70) \\\hline
$K^0_s$& 65(70) & 59(58) & 54(53) & 51(51)  \\\hline
\hline
\end{tabular}
\end{center}
\vskip-3mm
%\hspace*{17mm}\footnotesize{The error is of order of $1.5-2\%$ for all numbers
%in Tables~\ref{tab1} -- \ref{tab3}}
\end{table}

Though the neutral decay channels of the $\eta$ meson, like those of $K^0_s$ meson  (see
Table~\ref{tab:PDG}), have, on the average, four photons, the letter meson has noticeably less
discrimination powers \wrt single photons (especially with $\Et\gt40 ~GeV$) and 
from this point of view it is intermediate between $\eta$ and $\pi^0$ mesons. This is due to
a large difference (eight orders of magnitude) between the mean life times of the $\eta$ and $K^0_s$
mesons. In Table \ref{tab:ka_dec} we present a percentage of the decayed $K^0_s$
mesons up to the ECAL surface as a function of their $\Et$ and
pseudorapidity. Thus, as the energy increases the  $K^0_s$ decay
vertex becomes closer to the ECAL surface and for $\Et\gt 60 ~GeV$ 
the $\gamma/K^0_s$ and $\gamma/\pi^0$ discrimination powers are close enough
\footnote{see also Figs.~\ref{fig:rej_b0} -- \ref{fig:rej_e}}.
The same fact is reflected in Fig.~\ref{fig:NC80}, where
we plotted the normalized distribution of the number of events over the minimal number of crystal cells
containing $80\%$ of the ECAL tower energy for the initial particle ($\gamma, \pi^0, \eta, K^0_s$)
transverse energy $\Et= 40 ~GeV$. To find this number we summed 
cell energies $E$ in decreasing order starting with the most energetic cell
until the sum reached $80\%$ of the tower energy. The reason why the result for 
$K^0_s$ ($\left<Ncell\right>=2.9$) is intermediate between $\eta$ ($\left<Ncell\right>=3.7$) and
$\pi^0$ ($\left<Ncell\right>=2.6$) is given above.\\[-5mm]
\begin{table}[h]
%\noindent
\begin{center}
\vskip-4mm \caption{Percentage of the decayed $K^0_s$ mesons as a
function of their $\Et$ and pseudorapidity $\eta$. Only the
neutral decay channels are allowed.} 
\label{tab:ka_dec}
\vskip0.1cm
\begin{tabular}{||lc|c|c|c|c|c|c|c|} \hline
&          $\Et~~ (GeV)$ & 20 & 40 & 60 & 100 & 200 \\\hline\hline
&       $|\eta|\lt 0.4$  & 74.5 & 49.7 & 37.5 & 25.8 & 13.7 \\\hline
&$1.0\lt \eta\lt 1.4$  & 67.5 & 46.7 & 33.6 & 21.1 & 12.3 \\\hline
&$1.6\lt \eta\lt 2.4$  & 52.2 & 34.6 & 24.8 & 16.1 &  7.8 \\\hline
\hline
\end{tabular}
\end{center}
\end{table}
~\\[-10mm]

Bracketed figures in Tables~\ref{tab1} -- \ref{tab3} are the $\gamma/K^0_s$ (and
$\gamma/\eta$) separation probabilities which one would have if the neural network were trained with
the $\gamma/K^0_s$ (and $\gamma/\eta$) samples. We see that differences in separation 
probabilities by the ``0.5-criterion'' (for $K^0_s$ as an example) 
between the case that the network is {\it trained and tested} with the 
$\gamma/K^0_s$ event samples and the case that the network is {\it trained} with  $\gamma/\pi^0$ 
 {\it and tested} with $\gamma/K^0_s$ events sample are within about $4-8\%$ for all $\Et$ and 
pseudorapidity intervals considered.

For various applications it is useful to find which
rejection can be obtained for a given single photon selection
efficiency. The respective ``$\gamma$ selection/meson rejection''
curves are shown in Figs.~\ref{fig:rej_b0} -- \ref{fig:rej_e}
(also for three pseudorapidity $\eta$ intervals). The solid,
dotted and dashed lines  correspond to rejections of the
$\pi^0, K^0_s$ and $\eta$ meson multiphoton final states. The rejections are seen to 
gradually decrease with growing pseudorapidity.

As we mentioned in Section 4, in practice various network output
threshold values $O_{thr}$ can be used to achieve better 
signal-to-background ($S/B$) ratios at the cost of statistics loss. We varied
the output discriminator $O_{thr}$ value from 0.4 to 0.7. The
resulting $S/B$ (where $S$ corresponds to the single photon $\gamma$ events and
$B$ to the background neutral pion $\pi^0$ events) ratios for all 
$\Et$ values and $\eta$ intervals considered are given in Tables~\ref{tab:sb_b0} -- \ref{tab:sb_e}. 
One can see that the Signal/Background ratio grows with growing NN output threshold. 
For example, at $\Ptgp=20 ~GeV/c$ and for $|\eta|\lt0.4$ it grows
 from 2.67 to 6.30  and for the same pseudorapidity interval and at $\Ptgp=60 ~GeV/c$ 
it grows from 1.42 to 2.43 while $O_{thr}$ varies from 0.40 to 0.70. \\[-8mm]
%\newpage
\begin{table}[htbp]
\begin{center}
\caption{Signal($\gamma$)/Background($\pi^0$). $|\eta|\lt0.4$.}
\vskip.1cm
\begin{tabular}{||c||c|c|c|c|c|c|c||}                  \hline \hline
\label{tab:sb_b0}
$\Ptgp$  &\multicolumn{7}{c|}{NN output cut $O_{thr}$} \\\cline{2-8}
$(GeV/c)$& 0.40& 0.45& 0.50& 0.55& 0.60& 0.65& 0.70 \\\hline \hline
20       & 2.67& 3.06& 3.53& 4.07& 4.65& 5.43& 6.30   \\\hline
40       & 1.87& 2.13& 2.38& 2.60& 2.84& 3.11& 3.47   \\\hline
60       & 1.42& 1.50& 1.58& 1.71& 1.90& 2.15& 2.43  \\\hline
100      & 1.23& 1.27& 1.32& 1.42& 1.60& 1.73& 1.95  \\\hline
\hline
\end{tabular}
\end{center}
\vskip-14mm
\end{table}

\begin{table}[htbp]
\begin{center}
\vskip-4mm
\caption{Signal($\gamma$)/Background($\pi^0$). $1.0\lt\eta\lt1.4$.}
\vskip.2cm
\normalsize
\begin{tabular}{||c||c|c|c|c|c|c|c||}                  \hline \hline
\label{tab:sb_b1}
$\Ptgp$  &\multicolumn{7}{c|}{NN output cut $O_{thr}$} \\\cline{2-8}
$(GeV/c)$&0.40 &0.45 &0.50 &0.55 &0.60 &0.65 &0.70 \\\hline \hline
20       & 2.02& 2.38& 2.81& 3.37& 3.96& 4.65& 5.41 \\\hline
40       & 1.81& 2.03& 2.31& 2.54& 2.79& 3.02& 3.33 \\\hline
60       & 1.49& 1.59& 1.69& 1.93& 2.12& 2.28& 2.51  \\\hline
100      & 1.24& 1.51& 1.56& 1.63& 1.79& 1.97& 2.28  \\\hline
\hline
\end{tabular}
\end{center}
%\end{table}
%\begin{table}[htbp]
\begin{center}
\vskip1mm
\caption{Signal($\gamma$)/Background($\pi^0$). $1.6\lt\eta\lt2.4$.}
\vskip.1cm
\normalsize
\begin{tabular}{||c||c|c|c|c|c|c|c||}                \hline \hline
\label{tab:sb_e}
$\Ptgp$  &\multicolumn{7}{c|}{NN output cut $O_{thr}$} \\\cline{2-8}
$(GeV/c)$&0.40 &0.45 &0.50 &0.55 &0.60 &0.65 &0.70 \\\hline \hline
20       &1.31& 1.52& 1.79& 2.00& 2.42& 2.74& 3.44 \\\hline
40       &1.28& 1.32& 1.64& 1.83& 2.05& 2.24& 2.48\\\hline
60       &1.04& 1.05& 1.34& 1.56& 1.61& 1.84& --  \\\hline
\hline
\end{tabular}
\end{center}
\vskip-2mm
\hspace*{17mm}\footnotesize{The errors for $S/B$ values in Tables $7-9$ above 
are of order of 0.10-0.20.% for NN output less than 0.60 and greater than 0.60 - 0.65 respectively.
}
\end{table}

%=================================================================================
%
\section{Conclusion.}
%
%=================================================================================

One can make some concluding remarks on Figs.~\ref{fig:rej_b0} -- \ref{fig:rej_e} and
Tables~\ref{tab1} -- \ref{tab3}. 

For example, with $50\%$ and $75\%$ of single photon events, we obtain $S^{(\gamma)}/B^{(\pi^0)}$ 
ratios shown in Fig.~\ref{fig:sb_pt}.

The  results obtained here can be improved by additional
training of the network with the border patterns, i.e. with events for which the NN output $O$
is close to the $\gamma/\pi^0$ border value 0.5 for two classes of events. But it
would require at least 3-5 times larger statistics than considered here and,
consequently, huge computing resources.

The network performance appears to be very sensitive to the crystal cell size. The results obtained by
the authors in parallel with \cite{Bor} for the Barrel region with old ECAL
geometry (until 1998) with $0.0145\times0.0145$ cell size are
much better than those given in the present paper
\footnote{Taking $90\%$ of single photons at $\Et=40 ~GeV$, for example, 
one could reach the $\pi^0$ rejection efficiency equal to about $60\%$ for the old geometry
instead of about $30\%$ for the new one used here.}.
Not so impressive results of $\pi^0,\eta, K^0_s$ meson rejections obtained after analyzing Endcap
cells necessitates the use of a preshower in this region, perfect rejection powers for
which after the analysis based on the ANN application were shown in \cite{Bar2}.

\begin{center}
\begin{figure}[htbp]
\vskip-19mm
\hspace*{-10mm} \includegraphics[width=16cm,height=8.5cm,angle=0]{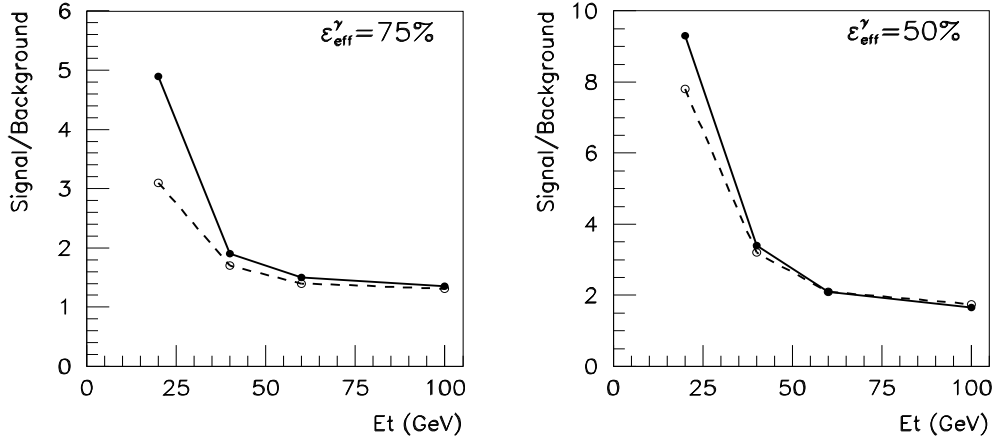}
\vskip-10mm
  \caption{Signal($\gamma$) to Background($\pi^0$) ratios over different $\Et$ values
for two values of photon selection efficiency
$\epsilon^{\gamma}_{eff}=50$ and $75\%$. The solid curves correspond to
the $|\eta|\lt0.4$ and dashed ones to the $1.0\lt|\eta|\lt1.4$
intervals.}
  \label{fig:sb_pt}
\end{figure}
\vskip-10mm
\end{center}
%So, we see from the plots in Fig.~\ref{fig:sb_pt}

\newpage
~\\[-1mm]

\noindent                                                            
{\bf Acknowledgements.}  \\
We are greatly thankful to G.~Gogiberidze (BNL) for the helpful discussions on 
the artificial neural network usage for pattern recognition tasks and 
to A.~Nikitenko (CERN) for supplying us with the digitization routines.

\begin{center}
\begin{figure}[htbp]
\vskip-1mm
\hspace*{-4mm} \includegraphics[width=15cm,height=16cm,angle=0]{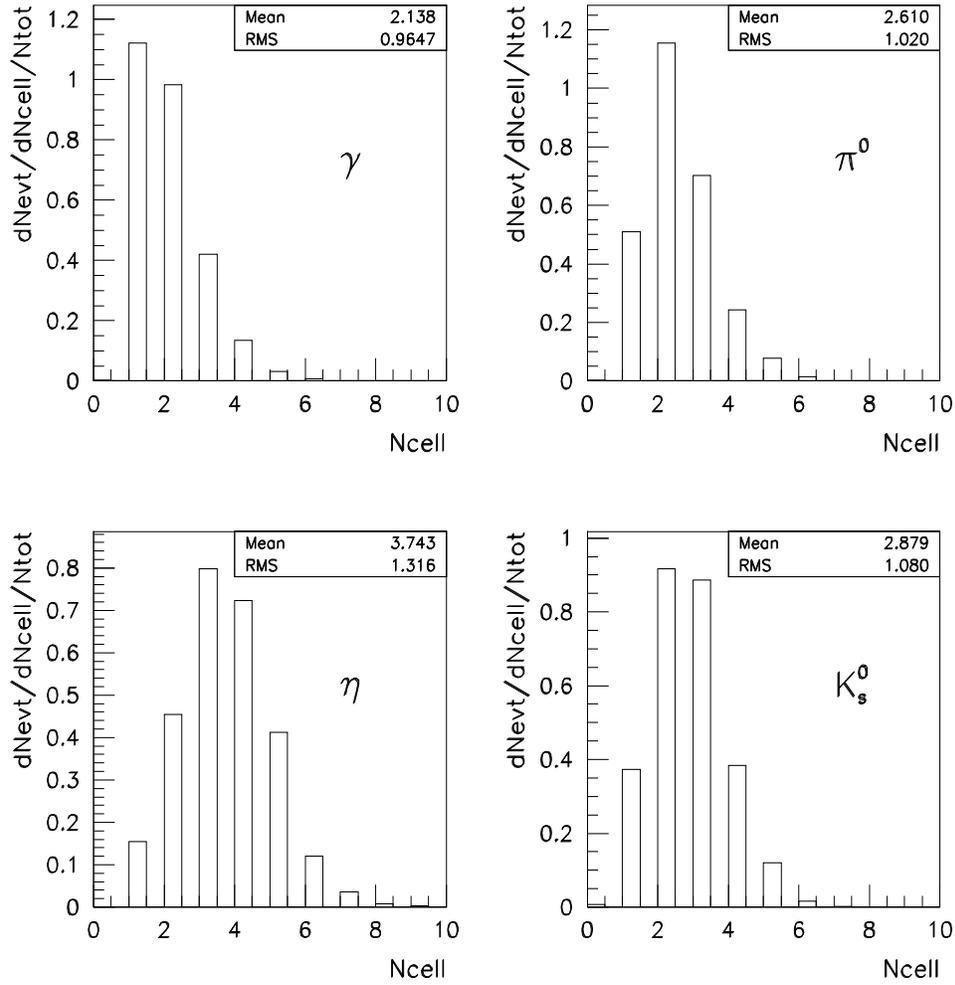}
\vskip-1mm
  \caption{Normalized distribution of events number over the minimal number of crystal cells
containing $80\%$ of ECAL tower energy. $\Et= 40 ~GeV$, $1.0\lt |\eta|\lt 1.4$.}
  \label{fig:NC80}
\end{figure}
%\end{center}
%\begin{center}
\begin{figure}[htbp]
\vskip-1mm
\hspace*{-4mm} \includegraphics[width=15cm,height=15cm,angle=0]{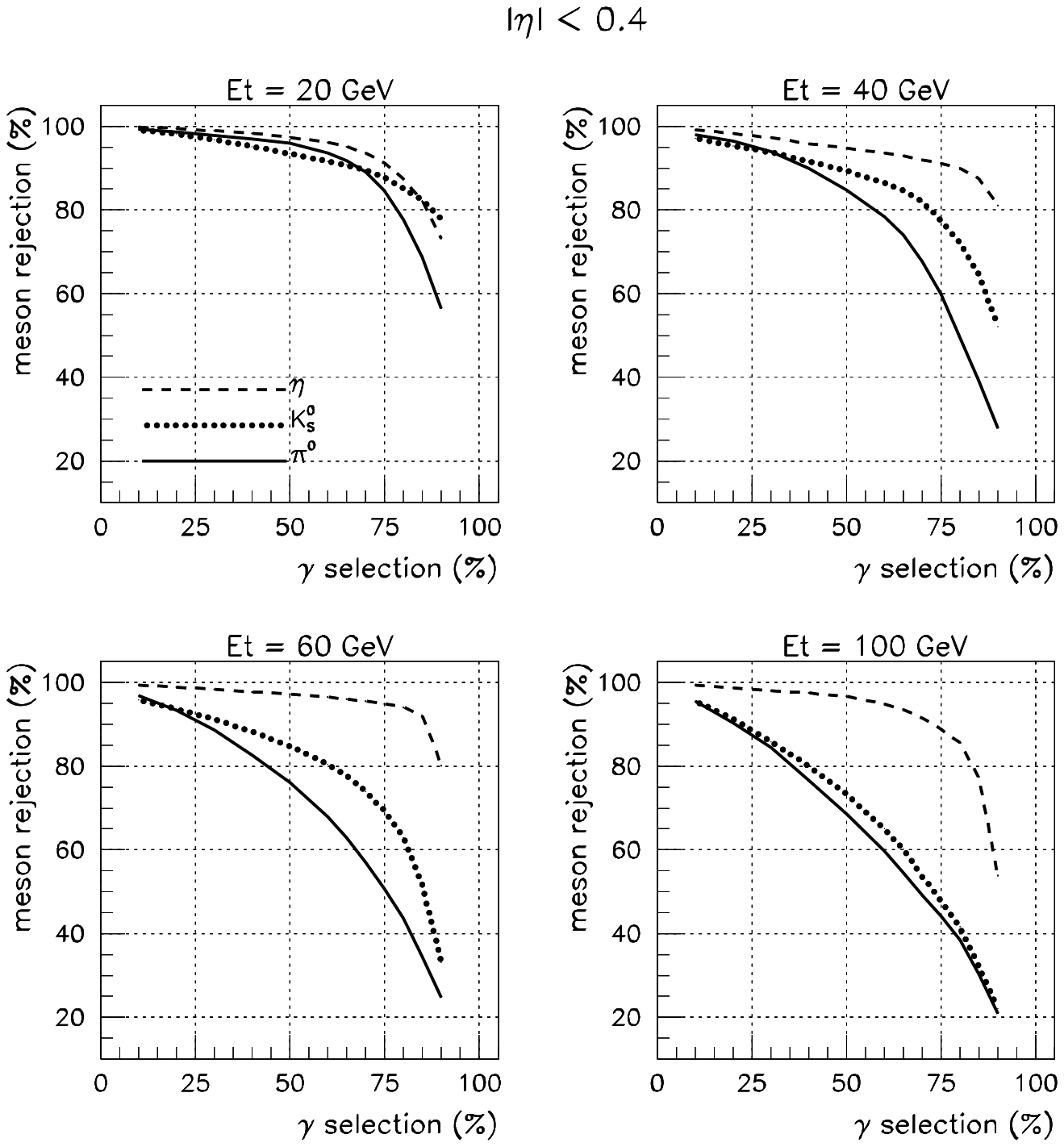}
\vskip-1mm
  \caption{Single photon selection efficiency vs. rejection of $\pi^0, \eta, K^0_s$ mesons for four
$\Et$ values and the $|\eta|\lt 0.4$ interval.}
  \label{fig:rej_b0}
\end{figure}
%\end{center}
%\begin{center}
\begin{figure}[htbp]
\vskip-1mm
\hspace*{-4mm} \includegraphics[width=15cm,height=15cm,angle=0]{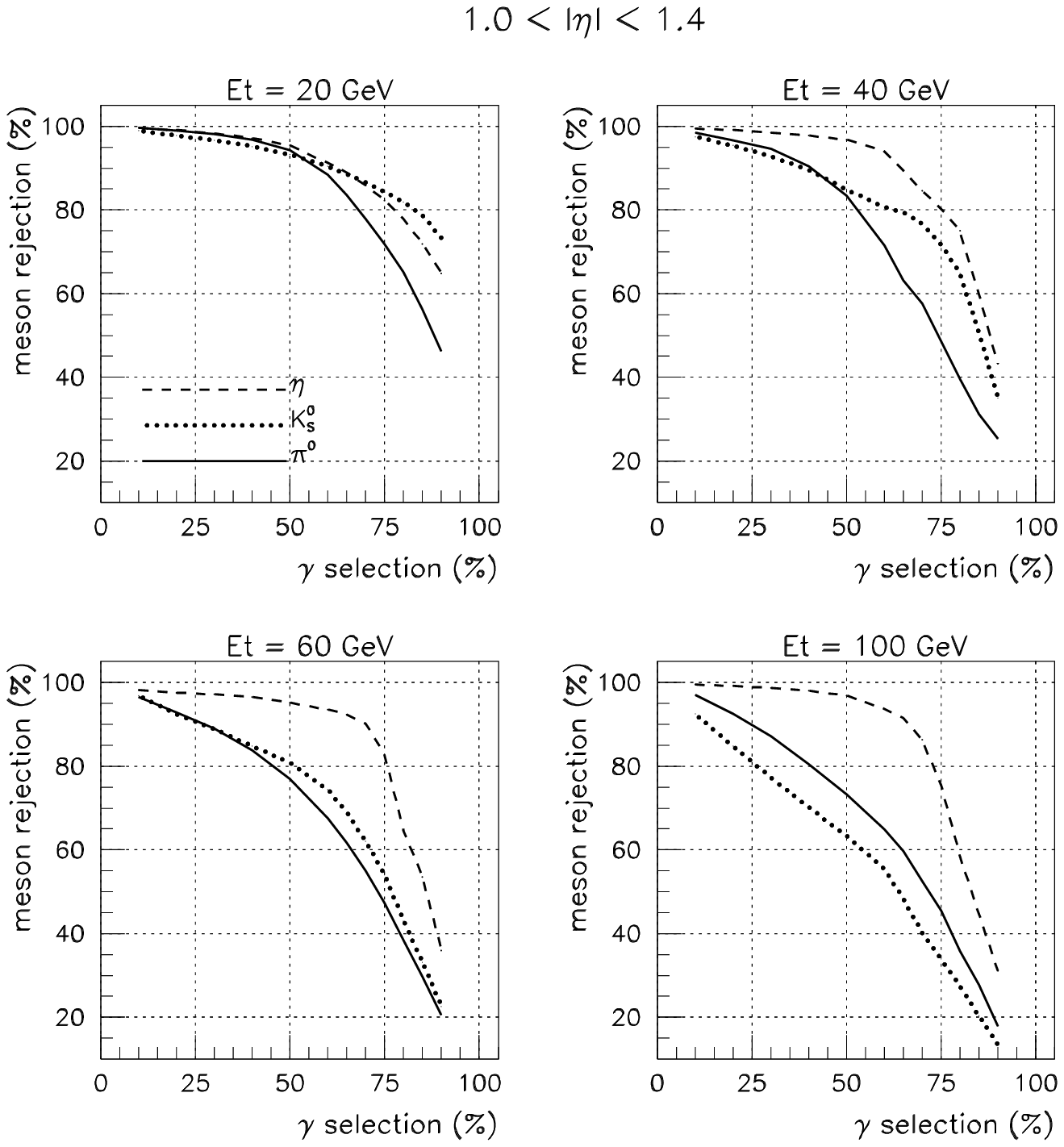}
\vskip-1mm
  \caption{Single photon selection efficiency vs. rejection of $\pi^0, \eta, K^0_s$ mesons for four
$\Et$ values and the $1.0\lt |\eta|\lt 1.4$ interval.}
  \label{fig:rej_b1}
\end{figure}
%\end{center}
%\begin{center}
\begin{figure}[htbp]
\vskip-1mm
\hspace*{-4mm} \includegraphics[width=15cm,height=15cm,angle=0]{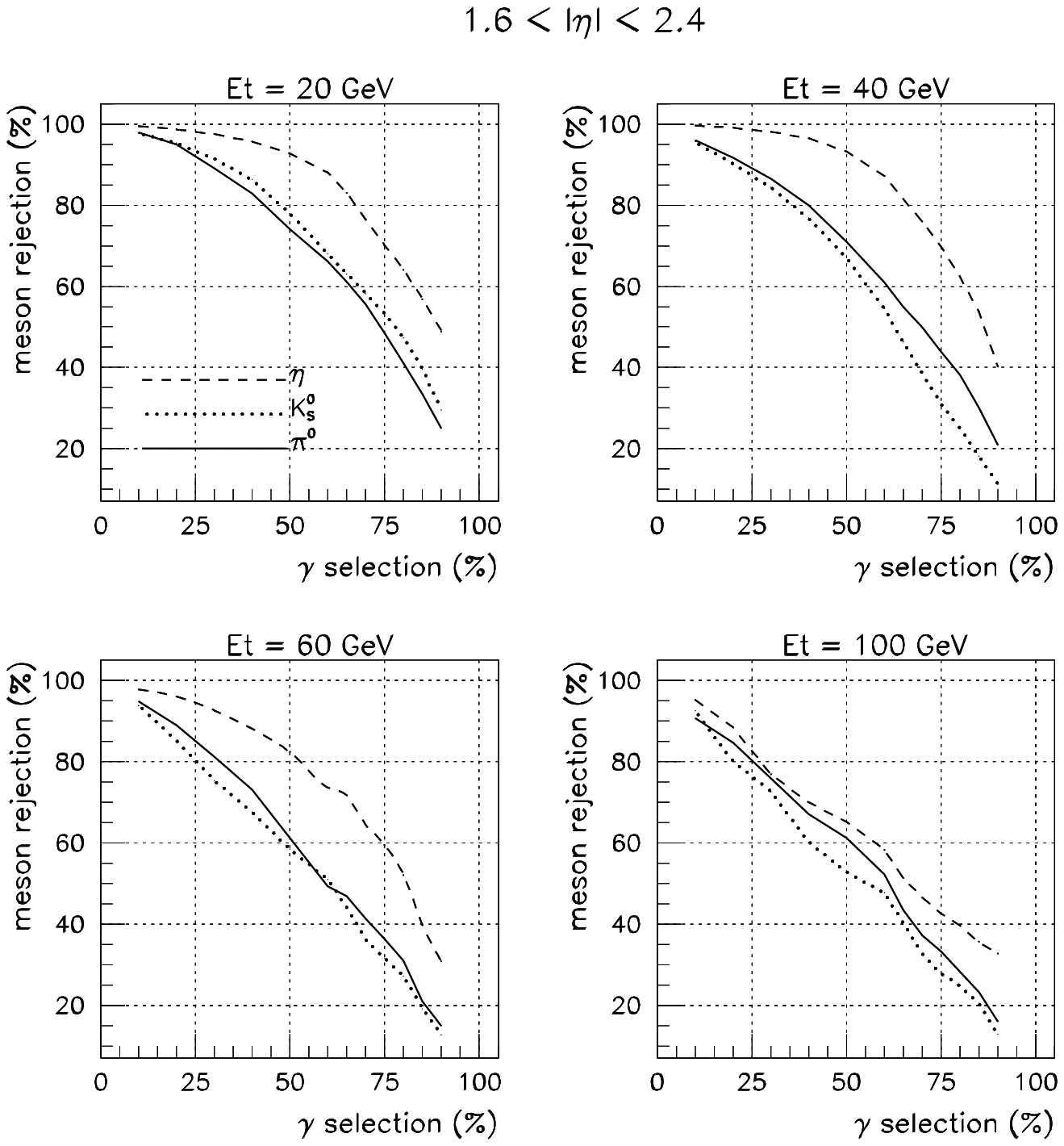}
\vskip-1mm
  \caption{Single photon selection efficiency vs. rejection of $\pi^0, \eta, K^0_s$ mesons for four
$\Et$ values and the $1.6 \lt |\eta|\lt 2.4$ interval.}
  \label{fig:rej_e}
\end{figure}
\end{center}


\begin{thebibliography}{99}

\bibitem{SF} %1
 D.V.~Bandourin, V.F.~Konoplyanikov, N.B.~Skachkov.
``$\gamma+Jet$ events rate estimation for gluon distribution determination at LHC'',
Part.Nucl.Lett.103:34-43, 2000, hep-ex 0011015.
\bibitem{BKS_P1} %4
 D.V.~Bandourin, V.F.~Konoplyanikov, N.B.~Skachkov.
``Jet energy scale setting with \gpj events at LHC energies.
Generalities, Selection rules'', JINR Preprint E2-2000-251, JINR, Dubna,
hep-ex 0011012.
\bibitem{BKS_P5} %4
 D.V.~Bandourin, V.F.~Konoplyanikov, N.B.~Skachkov.
``Jet energy scale setting with \gpj events at LHC energies.
Detailed study of the background suppression.'', JINR Preprint E2-2000-255,
JINR, Dubna, hep-ex 0011017.
\bibitem{Bar1}
D.~Barney, P.~Bloch, CMS-TN/95-114,
``$\pi^0$ rejection in the CMS endcap electromagnetic calorimeter - with and
without a preshower.'
\bibitem{Bar2}
A.~Kyriakis, D.~Loukas, J.~Mousa, D.~Barney, CMS Note 1998/088,
``Artificial neural net approach to $\gamma-\pi^0$ discrimination using CMS Endcap
Preshower''.
\bibitem{Bor}
L.~Borissov, A.~Kirkby, H.~Newman, S.~Shevchenko, CMS Note 1997/050,
``Neutral pion rejection in the CMS $PbWO_4$ crystal calorimeter using a
neural network''.
\bibitem{CMSIM}
GEANT-3 based simulation package of CMS detector, CMSIM, Version 116.
CMS TN/93-63, C.~Charlot {\it et al}, ``CMSIM--CMANA. CMS Simulation facilities'',
CMSIM User's Guide at WWW: http://cmsdoc.cern.ch/cmsim/cmsim.html.
\bibitem{PDG} Particle Data Group,
D.E. Groom {\it et al.}, The European Physical Journal C15 (2000) 1.
\bibitem{JN} %2
C.~Peterson, T.~Rognvaldsson and L~Lonnblad,
``JETNET 3.0. A versatile Artificial Neural Network Package'',
Lund University Preprint LU-TP 93-29.

\end{thebibliography}
\end{document}